\documentclass[aps,pre,twocolumn,superscriptaddress]{revtex4-1}

\usepackage{graphicx}
\usepackage{dcolumn}
\usepackage{bm}

\newcommand{\beq}{\begin{equation}}
\newcommand{\eeq}{\end{equation}}
\newcommand{\beqa}{\begin{eqnarray}}
\newcommand{\eeqa}{\end{eqnarray}}
\newcommand{\Tr}{\text{Tr}}

\newcommand{\bra}[1]{\left \langle #1 \right |}
\newcommand{\ket}[1]{\left | #1 \right \rangle}
\newcommand{\av}[1]{\left\langle #1 \right\rangle}

\newcommand{\kket}[1]{| #1 \hspace{-.9mm}\left. \right>}
\newcommand{\bbra}[1]{  \hspace{-.6mm}\left<\hspace{-.4mm} \right. #1 |  }
\newcommand{\aav}[1]{ \hspace{-.6mm}\left<\hspace{-.4mm} \right. #1 \hspace{-.9mm}\left. \right> }

\newcommand{\ii}{\mathrm{i}}
\newcommand{\ee}{\mathrm{e}}

\newcommand{\bz}{k_{\text{B}}}

\begin{document}


\title{Integral Quantum Fluctuation Theorems under Measurement and Feedback control}


\author{Ken Funo}
\affiliation{Department of Physics, The University of Tokyo, 7-3-1 Hongo, Bunkyo-ku, Tokyo 113-0033, Japan}
\author{Yu Watanabe} 
\affiliation{Yukawa Institute for Theoretical Physics, Kyoto University, Kitashirakawa Oiwake-Cho, Kyoto 606-8502, Japan}
\author{Masahito Ueda}
\affiliation{Department of Physics, The University of Tokyo, 7-3-1 Hongo, Bunkyo-ku, Tokyo 113-0033, Japan}

\date{\today}

\begin{abstract}
We derive integral quantum fluctuation theorems and quantum Jarzynski equalities for a feedback-controlled system and a memory which registers outcomes of the measurement. The obtained equalities involve the information content, which reflects the information exchange between the system and the memory, and take into account the back action of a general measurement contrary to the classical case. The generalized second law of thermodynamics under measurement and feedback control are reproduced from these equalities.
\end{abstract}


\maketitle


\section{\label{sec:Intro}Introduction}

Fluctuation theorems have attracted considerable interest both theoretically~\cite{Crooks,Evans,Gallavotti} and experimentally~\cite{Jarexperiment,Flucex}, since they lead to several fundamental relations in nonequilibrium statistical mechanics. In particular, the Jarzynski equality~\cite{Jarzynski1,Jarzynski2} allows us to relate the nonequilibrium work to the equilibrium free-energy difference, and it was used to find the free-energy difference of a single molecule from the measured work~\cite{Jarexperiment,Flucex}. Fluctuation theorems and various Jarzynski equalities for quantum systems have been explored~\cite{fluctuation1,Tasaki,Kurchan,characteristic1,characteristic2,Esposito,Horowitz1,Horowitz2,Monnai,Campisi,Chetrite,Hu,Crooks2}. An experiment on a double quantum-dot system was done in Ref.~\cite{Utsumi}, verifying the fluctuation theorem concerning the probability distribution of the number of electrons tunneling through a double quantum dot. An experimental reconstruction of the work distribution function and the verification of the quantum Jarzynski equality were done in an NMR system~\cite{Flucexp}. The integral fluctuation theorem without feedback control is given by
\beq
\aav{\ee^{-\sigma}}=1, \label{introfluc}
\eeq
where $\sigma$ is the entropy production, which is related to the equilibrium free-energy difference $\Delta F$ and the work $W$ done by the system as $\sigma=-\beta(W+\Delta F)$, where the initial state is assumed to be the canonical distribution with inverse temperature $\beta$. By using the Jensen inequality $\aav{\ee^{-\sigma}}\geq \ee^{-\aav{\sigma}}$, and noting that $\ee^{-\aav{\sigma}}\geq 1-\aav{\sigma}$, Eq.~(\ref{introfluc}) reproduces the second law of thermodynamics:
\beq
\aav{\sigma}\geq 0.
\eeq
As pointed out by Maxwell~\cite{JMaxwell}, if a feedback controller (Maxwell's demon) can access the microscopic degrees of freedom, the second law of thermodynamics may be modified. Szilard explicitly constructed a model in which a feedback controller can extract the work of $W=\bz T\ln 2$ from the system during a thermodynamic cycle by utilizing one bit (=$\ln2$) of information gain obtained by the measurement~\cite{Szilard}. By using a correlation between the system and the memory which registers the measurement outcome, the feedback controller can decrease the entropy of a small fluctuating system at the cost of increasing the entropy of the memory. It has been shown that the work gain from the system due to the feedback control is characterized by the mutual information content which expresses the correlation between the system and the memory, and also sets a lower bound on the information processing cost~\cite{Sagawa1,Sagawa2,Sagawa3,Sagawa4}. The second law of thermodynamics under information processing is given by~\cite{Sagawa3}
\beq
\aav{\sigma^{S}}\geq \Delta I, \ \aav{\sigma^{M}}\geq \Delta I' \label{generalizedfluctuationent}
\eeq
where $\sigma^{S(M)}:=\Delta S^{S(M)}-\beta Q^{S(M)}$ with $\Delta S^{S(M)}$ being a change in the Shannon entropy of the system (memory) and $Q^{S(M)}$ being the heat transfered from the heat bath to the system (memory), and $\Delta I$, $\Delta I'$ are the changes in the classical mutual information between the system and the memory via information processing. Note that the second law of thermodynamics dictates that the entropy production be nonnegative. In contrast, the right-hand side of Eq.~(\ref{generalizedfluctuationent}) can be negative, since $\sigma^{S}$ and $\sigma^{M}$ contain not only the dissipative part but also the information exchange between the system and the memory. Similar inequalities can be obtained for quantum systems if there is neither an initial nor a final correlation between the system and the memory~\cite{Sagawa1,Sagawa2,Sagawa4,Sagawa5}~: $\aav{\sigma^{S}}\geq -I$ and $\aav{\sigma^{M}}\geq +I$, where $I$ is the information gain on the system~\cite{Groenewold,Ozawa}.

Associated with this identification of the information-theoretic quantity in the second law of thermodynamics under information processing, we may expect that the same quantity should also play a crucial role in fluctuation theorems. In the classical case, fluctuation theorems have been generalized by including the mutual information content or introducing the efficacy parameter $\gamma$ of feedback control~\cite{SagawaJar}:
\beqa
\aav{\ee^{-\sigma^{S}+\Delta I}}&=&1,\ \aav{\ee^{-\sigma^{M}+\Delta I'}}=1,  \label{classicalfluc} \\
\aav{\ee^{-\sigma^{S}}}&=&\gamma. \label{gamma}
\eeqa
An experimental demonstration of Maxwell's demon was carried out by performing a feedback control on a Brownian particle in a tilted washboard potential~\cite{Toyabe}, and the fluctuation theorem in the form of Eq.~(\ref{gamma}) has been verified. In Ref.~\cite{Morikuni}, the authors have obtained quantum versions of Eqs.~(\ref{classicalfluc}) and~(\ref{gamma}), the former being given by
\beq
\aav{\ee^{-\sigma^{S}-I}}=1,\  \sigma^{S}=-\beta(W^{S}+\Delta F^{S}).  \label{Morikuni}
\eeq
Here, they consider a projection measurement on the system followed by a process with classical error, so that the outcome $j'$ may differ from the actual postmeasurement state $\ket{j}$ of the system according to the error probability $\epsilon(j\rightarrow j')$.

In this paper, we derive quantum fluctuation theorems under general quantum measurement and feedback control by taking into account the measurement back action within the framework of operational quantum measurement theory~\cite{Kraus,Nielsen}, and show that the obtained fluctuation theorems involve the information gain which characterizes the acquired knowledge about the system due to quantum measurement. The obtained equalities reproduce the second law of thermodynamics under measurement and feedback control obtained in Refs.~\cite{Sagawa1,Sagawa2}.

This paper is organized as follows. In Sec.~\ref{sec:Formulation}, we introduce a protocol to realize measurement and feedback control. In Sec.~\ref{sec:Main}, we derive integral quantum fluctuation theorems under the protocol discussed in Sec.~\ref{sec:Formulation}. In Sec.~\ref{sec:summary}, we summarize the main results of this paper.


\section{\label{sec:Formulation}Formulation of the problem}

\begin{figure*}[t]
\includegraphics[width=.95\textwidth]{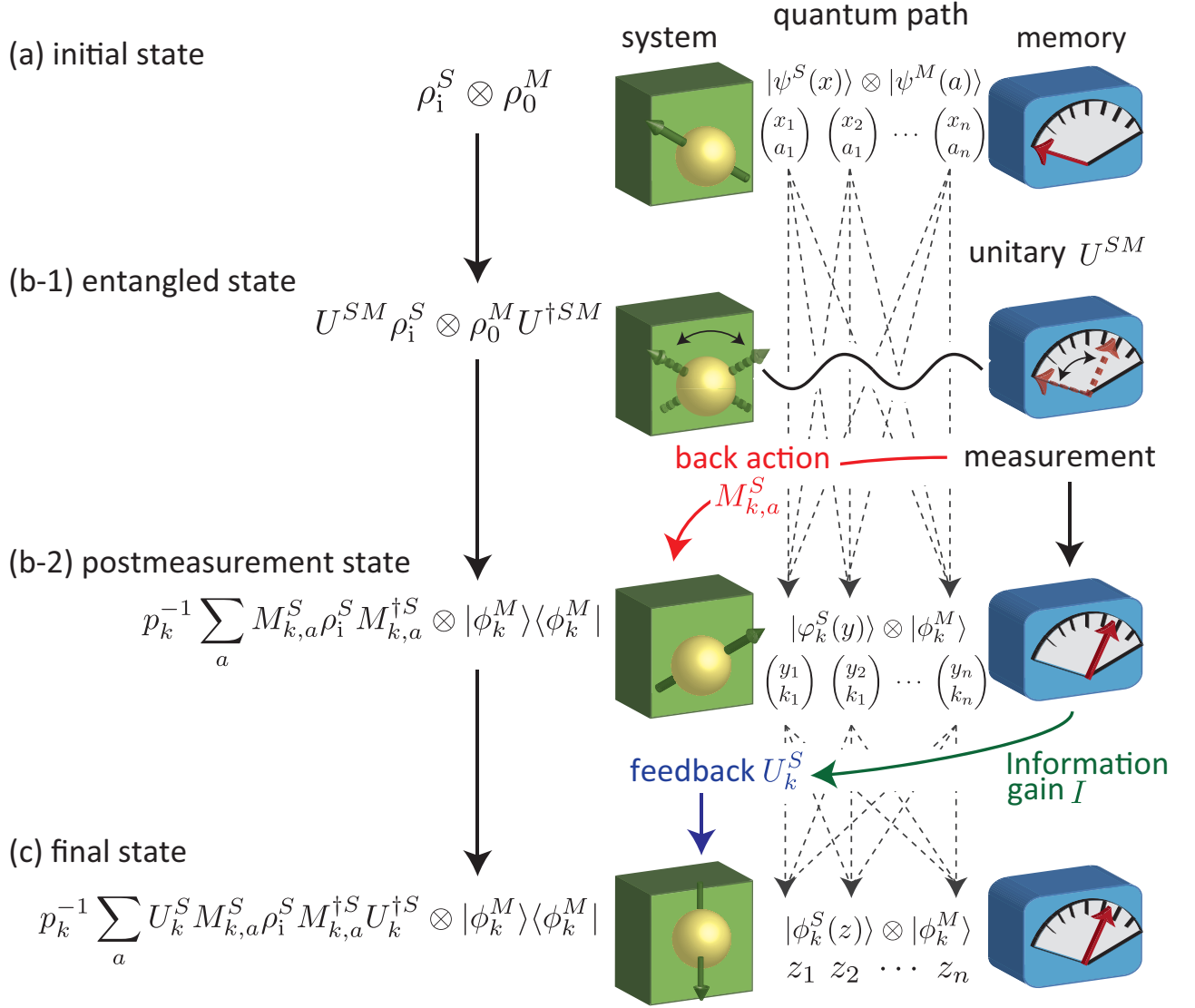}
\caption{\label{fig1} (Color online) Schematic illustration of our setup. (a) The initial state of the system and memory is given by $\rho^{S}_{\mathrm{i}}\otimes\rho^{M}_{0}$, and its diagonal basis is given by $\kket{\psi^{S}(x)}\otimes\kket{\psi^{M}(a)}$. (b-1) A unitary transformation $U^{SM}$ is performed to entangle the system and memory. After this transformation, the state is given by $U^{SM}\rho^{S}_{\mathrm{i}}\otimes\rho^{M}_{0}U^{\dagger SM}$. (b-2) A general measurement $M_{k,a}:=\sqrt{p^{M}_{a}}\bbra{\phi^{M}_{k}}U^{SM}\kket{\phi^{M}_{a}}$ on the system is performed through a projection measurement on the memory by using the basis $\kket{\phi^{M}_{k}}$, and the measurement outcome $k$ is obtained. The postmeasurement state is given by $p^{-1}_{k}\sum_{a}M^{S}_{k,a}\rho^{S}_{\mathrm{i}}M^{\dagger S}_{k,a}\otimes\kket{\phi^{M}_{k}}\bbra{\phi^{M}_{k}}$, and its diagonal basis is given by $\kket{\varphi^{S}_{k}(y)}\otimes\kket{\phi^{M}_{k}}$. The measurement process gives a quantum transition between states labeled by $(x,a)$ and $(y,k)$. (c) From the measurement outcome $k$, we acquire information about the system, which is characterized by the information gain $I$. It is utilized to perform a feedback control on the system described by a unitary transformation $U^{S}_{k}$, which depends on the measurement outcome $k$. The final state is given by $\rho^{SM}_{\mathrm{f}}=p^{-1}_{k}\sum_{a}U^{S}_{k}M^{S}_{k,a}\rho^{S}_{\mathrm{i}}M^{\dagger S}_{k,a}U^{\dagger S}_{k}\otimes\kket{\phi^{M}_{k}}\bbra{\phi^{M}_{k}}$. The entropy production or work can be evaluated from the matrix element $\bbra{\phi^{S}_{k}(z)}\rho^{S}_{\mathrm{f}}(k)\kket{\phi^{S}_{k}(z)}$, where $\kket{\phi^{S}_{k}(z)}$ is the diagonal element of the reference state or the energy eigenstate of the final Hamiltonian. Thus, a feedback control gives a quantum transition between the states labeled by $(y,k)$ and $z$.}
\end{figure*}

We consider a general measurement and feedback control by introducing a memory which registers measurement outcomes and assume that the system and memory are isolated from the heat bath. The effect of heat bath on the quantum fluctuation theorem has been studied in Refs.~\cite{Chetrite,Hu,Crooks2,fluctuation1,Horowitz1,Horowitz2}. We follow the following procedures to perform measurement and feedback (see Fig.~\ref{fig1}).

(a) Let $\rho^{S}_{\mathrm{i}}$ and $\rho^{M}_{\mathrm{0}}$ be the initial states of the system and the memory, respectively.

(b) A general quantum measurement on the system is implemented by performing a unitary transformation $U^{SM}$ on the composite system followed by a projection measurement $P^{M}_{k}:=\kket{\phi^{M}_{k}}\bbra{\phi^{M}_{k}}$ on the memory. The postmeasurement state depending on the measurement outcome $k$ is given by
\beqa
\rho^{SM}(k)&=&\frac{1}{p_{k}} P^{M}_{k} U^{SM}(\rho_{\mathrm{i}}^{S}\otimes\rho^{M}_{0})U^{\dagger SM} P^{M}_{k} \nonumber \\
&=&\sum_{a}\frac{M^{S}_{k,a}\rho^{S}_{\mathrm{i}}M^{\dagger S}_{k,a}}{p_{k}}\otimes\kket{\phi^{M}_{k}}\bbra{\phi^{M}_{k}}, \label{poststateabk}
\eeqa
where $M^{S}_{k,a}:=\sqrt{p^{M}_{0}(a)}\bbra{\phi^{M}_{k}}U^{SM}\kket{\psi^{M}(a)}$ is the measurement operator satisfying the relation $\sum_{k,a}M^{\dagger S}_{k,a}M^{S}_{k,a}=1$, and $\kket{\psi^{M}(a)}$ and $p^{M}_{0}(a)$ are given by the spectral decomposition of the initial state of the memory: $\rho_{0}^{M}=\sum_{a}p^{M}_{0}(a)\kket{\psi^{M}(a)}\bbra{\psi^{M}(a)}$. In Eq.~(\ref{poststateabk}), $p_{k}:=\Tr_{SM}[P^{M}_{k} U^{SM}(\rho_{\mathrm{i}}^{S}\otimes\rho^{M}_{0})U^{\dagger SM} P^{M}_{k}]$ is the probability of obtaining the outcome $k$.

(c) We perform a unitary transformation $U^{S}_{k}$ depending on the measurement outcome $k$. Here the unitary transformation is given by $U^{S}_{k}:=\text{T}\exp[-\ii\int^{t_{\mathrm{f}}}_{0}H_{k}^{S}(t)\mathrm{d}t]$, where $\text{T}$ is the time-ordering operator and $H^{S}_{k}(t)$ is the Hamiltonian of the system at time $t$ depending on the measurement outcome $k$. We note that the above unitary operation associated with the measurement outcome is nothing but the feedback control. The final state is given by
\beq
\rho^{SM}_{\mathrm{f}}(k)=U^{S}_{k}\rho^{SM}(k) U^{\dagger S}_{k}.
\eeq

Next, let us reconsider the above argument by expressing the initial state and the postmeasurement state in the diagonal basis.

(a') The spectral decompositions of the initial states of the system and the memory are given by $\rho^{S}_{\mathrm{i}}=\sum_{x}p^{S}_{\mathrm{i}}(x)\kket{\psi^{S}(x)}\bbra{\psi^{S}(x)}$ and $\rho_{0}^{M}=\sum_{a}p^{M}_{0}(a)\kket{\psi^{M}(a)}\bbra{\psi^{M}(a)}$. Now each element of the diagonal basis is labeled by the set of variables $(x,a)$. 

(b') We also decompose the postmeasurement state as follows: 
\beq
\rho^{SM}(k)=\sum_{y}p(y|k)\kket{\varphi^{S}_{k}(y)}\bbra{\varphi^{S}_{k}(y)}\otimes\kket{\phi^{M}_{k}}\bbra{\phi^{M}_{k}}. \label{diagpost}
\eeq
The measurement process gives a quantum transition from $\kket{\psi^{S}(x)}\otimes\kket{\psi^{M}(a)}$ to $\kket{\varphi^{S}_{k}(y)}\otimes\kket{\phi^{M}_{k}}$  characterized by the set of labels $(x,a,k,y)$. 

(c') The characterization of the quantum path that ends with the final state needs a little discussion. The final energy of the system is given by $E^{S}_{\mathrm{f}}(k):=\Tr[\rho^{S}_{\mathrm{f}}(k)H^{S}_{\mathrm{f}}(k)]=\sum_{z}E^{S}_{\mathrm{f}}(z|k)\bbra{\phi^{S}_{k}(z)}\rho^{S}_{\mathrm{f}}(k)\kket{\phi^{S}_{k}(z)}$, where $H^{S}_{\mathrm{f}}(k)$ is the final Hamiltonian of the system, $\{\kket{\phi^{S}_{k}(z)}\}$ is the set of energy eigenstates and $\{E^{S}_{\mathrm{f}}(z|k)\}$ is the set of the corresponding energy eigenvalues. Thus, we need to consider a quantum transition that ends with the state $\kket{\phi^{S}_{k}(z)}$ and connects the basis labeled by $(k,y)$ and $z$ during the feedback process.

The joint probability distribution of observing such a quantum transition connecting the basis labeled by $x,a,k,y$ and $z$ is given by
\beq
p(x,a,k,y,z):= p^{S}_{\mathrm{i}}(x)p^{M}_{0}(a)p(k,y|x,a)p(z|k,y), \label{totalprob}
\eeq
where
\beqa
p(k,y|x,a)&:=&|  \bbra{ \varphi^{S}_{k}(y)}\otimes\bbra{\phi^{M}_{k}}  U^{SM} \kket{ \psi^{S}(x) }\otimes\kket{\psi^{M}(a)} |^{2} \nonumber \\
&=&\frac{1}{p^{M}_{0}(a)}|  \bbra{ \varphi^{S}_{k}(y)}M^{S}_{k,a}\kket{ \psi^{S}(x) }|^{2}
\eeqa
is the conditional probability of obtaining $k$ and $y$ conditioned on $x$ and $a$, and
\beq
p(z|k,y):=|\bbra{\phi^{S}_{k}(z)}U^{S}_{k}\kket{\varphi^{S}_{k}(y)}|^{2}
\eeq
is the conditional probability of obtaining $z$ conditioned on $k$ and $y$. If we wish to obtain these probability distributions explicitly, we need to perform projective measurements on the system and memory~\cite{fluctuation1,Tasaki,Kurchan,characteristic1,characteristic2,Esposito,Horowitz1,Horowitz2,Monnai,Campisi}.

For each quantum path, we define physical quantities such as the total entropy production and the information gain as follows: The (unaveraged) total entropy production $\sigma^{SM}$ of the composite system is defined as
\beq
\sigma^{SM}(x,a,k,z):=\ln [p^{S}_{\mathrm{i}}(x)p^{M}_{0}(a)] -\ln [p^{S}_{\mathrm{r}}(z|k)p^{M}_{\mathrm{r}}(k)] \label{EPSM}.
\eeq
Here $p^{S}_{\mathrm{r}}(z|k)$ and $p^{M}_{\mathrm{r}}(k)$ are the reference states, each of which is related to the initial state of the system and memory in the backward process~\cite{Sagawa5}. Fluctuation theorems can be derived for any choice of these reference states; in the present paper, we assume that reference states are given by the canonical distributions corresponding to the final energy eigenvalues when deriving the quantum Jarzynski equality, i.e., $p^{S}_{\mathrm{r}}(z|k)=Z_{k}^{-1}\ee^{-\beta E^{S}_{\mathrm{f}}(z|k)}$, where $Z_{k}$ is the normalization constant. The averaged value of the total entropy production $\sigma^{SM}$ is nonnegative:
\beqa
\aav{\sigma^{SM}}&:=&\sum_{x,a,k,y,z}p(x,a,k,y,z)\sigma^{SM}(x,a,k,z) \nonumber \\
&=&\sum_{x,a,k,y,z}p(x,a,k,y,z)\ln \frac{p(x,a,k,y,z)}{\tilde{p}(z,y,k,x,a)} \nonumber\\
&\geq&0 , \label{totalslaw}
\eeqa
where the last inequality results from the positivity of the relative entropy (Kullback Leibler divergence)~\cite{Sagawa5,Nielsen}, and
\beq
\tilde{p}(z,y,k,x,a):=p_{\mathrm{r}}^{S}(z|k)p^{M}_{\mathrm{r}}(k)p(z|k,y)p(k,y|x,a)
\eeq
is the joint probability distribution of obtaining outcomes $z,y,k,x,a$ in the backward process.

Since we are interested in the effect of measurement and feedback on subsystems, we wish to decompose the total entropy production~(\ref{EPSM}) into the system and memory. For this purpose, let us consider the following decompositions:
\beqa
\sigma^{S}(x,k,z)&:=& \ln p^{S}_{\mathrm{i}}(x) -\ln p^{S}_{\mathrm{r}}(z|k), \label{EPS} \\
\sigma^{M}(a,k)&:=&\ln p^{M}_{0}(a)- \ln p^{M}_{\mathrm{r}}(k), \label{EPM}
\eeqa
where $\sigma^{SM}(x,a,k,z)=\sigma^{S}(x,k,z)+\sigma^{M}(a,k)$ is satisfied. The definitions of Eq.~(\ref{EPS}) is chosen to meet with the notation used in Ref.~\cite{Sagawa5}. As we discuss later, the averaged value of Eqs.~(\ref{EPS}) and (\ref{EPM}) can be negative since it contains not only the dissipative term but also the effect of information exchange between the system and memory. Such an effect of the information exchange can be expressed by the (unaveraged) information gain $I$ which is defined as: 
\beq
I(x,k,y)=\ln p(y|k)- \ln p^{S}_{\mathrm{i}}(x) ,  \label{genejarqcmutualdef}
\eeq
where $p(y|k)$ is the diagonal element of the postmeasurement state defined in Eq.~(\ref{diagpost}), and the averaged value of $I$ is equal to the information gain~\cite{Ozawa,Groenewold}. We will also discuss later that the combinations $\sigma^{S}+I$ and $\sigma^{M}-I$ measure the true entropy productions during the measurement and feedback processes. The information content that arises in the classical system is given by $I=\ln p(x|k)-\ln p_{\mathrm{i}}^{S}(x)$, where $p(x|k)$ is the conditional probability distribution of the system being $x$ conditioned on the measurement outcome $k$. On the other hand, Eq.~(\ref{genejarqcmutualdef}) depends on $y$ because the measurement back action alters the postmeasurement state. We will use these results to derive quantum fluctuation theorems and Jarzynski equalities under quantum measurement and feedback control.

%

\section{\label{sec:Main}Quantum fluctuation theorems and Jarzynski equalities}

\subsection{\label{sec:QFT}Quantum fluctuation theorems}

We show the following generalized quantum fluctuation theorems for the system and the memory:
\beq
\av{\ee^{-\sigma^{S}-I}}=1,  \label{QCfluc}
\eeq
\beq
\av{\ee^{-\sigma^{M}+I}}=1. \label{geneflucmemory}
\eeq
We assume that 
\beq
p(y|k)\neq 0 \hspace{5mm} \text{ for all } y \label{assumption}
\eeq
in deriving Eq.~(\ref{QCfluc}). In deriving Eq.~(\ref{geneflucmemory}), we assume that the initial probability distributions are non-vanishing, i.e.,
\beq
p^{S}_{\mathrm{i}}(x)\neq 0, \hspace{5mm} p^{M}_{0}(a)\neq 0\hspace{5mm} \text{for all }x \text{ and }a. \label{assumptiona}
\eeq
At the end of this subsection, we consider the case for which assumptions~(\ref{assumption}) and~(\ref{assumptiona}) do not hold. It follows from these assumptions that $p(k,y|x,a)$ and $p(z|k,y)$ are doubly stochastic:
\beq
\sum_{x,a}p(k,y|x,a)=\sum_{k,y}p(k,y|x,a)=1, \label{doublystochasticxa}
\eeq
which is obtained from the unitarity of $U^{SM}$. We also have
\beq
\sum_{y}p(z|k,y)=\sum_{z}p(z|k,y)=1, \label{doublystochasticz}
\eeq
which is obtained from the unitarity of $U^{S}_{k}$. By explicitly calculating the left-hand side of Eq.~(\ref{QCfluc}), we have
\beqa
\av{\ee^{-\sigma^{S} - I }}&:=& \sum_{x,a,k,y,z}p(x,a,k,y,z)\ee^{-\sigma^{S}(x,k,z)-I(x,k,y)} \nonumber \\
&=&  \sum_{x,a,k,y,z} p(x,a,k,y,z)\frac{p^{S}_{\mathrm{r}}(z|k)}{p(y|k)} \nonumber \\
&=& \sum_{k,y,z}p_{k} p^{S}_{\mathrm{r}}(z|k) p(z|k,y)=1, \label{proofents}
\eeqa
where the third equality follows from
\beqa
& &\sum_{x,a}p(x,a,k,y,z) \nonumber \\
&=&\bbra{\varphi^{S}_{k}(y)}\otimes\bbra{\phi^{M}_{k}}p_{k}\rho^{SM}(k)\kket{\varphi^{S}_{k}(y)}\otimes\kket{\phi^{M}_{k}} p(z|k,y) \nonumber \\
&=&p_{k}p(y|k)p(z|k,y),
\eeqa
and the last equality in Eq.~(\ref{proofents}) results from Eq.~(\ref{doublystochasticz}). Similarly, for the memory, we have
\beqa
\av{ \ee^{-\sigma^{M}+I}}&:=&\sum_{x,a,k,y,z}p(x,a,k,y,z)\ee^{-\sigma^{M}(a,k)+I(x,k,y)}     \nonumber \\
&=&\sum_{x,a,k,y}p^{M}_{\mathrm{r}}(k)p(y|k) p(k,y|x,a)  \nonumber \\
&=&\sum_{k,y}p^{M}_{\mathrm{r}}(k)p(y|k) =1,
\eeqa
where the third equality follows from Eq.~(\ref{doublystochasticxa}).

Now we derive the second law of thermodynamics under feedback control using Eqs.~(\ref{QCfluc}) and (\ref{geneflucmemory}). From Eq.~(\ref{QCfluc}), the Jensen inequality $\aav{\ee^{x}}\geq\ee^{\av{x}}$ and the relation $\ee^{\av{x}}\geq 1+\av{x}$, we obtain
\beq
\aav{\sigma^{S}}\geq -\av{I}. \label{sepslaw}
\eeq
Similarly, it follows from Eq.~(\ref{geneflucmemory}) that
\beq
\aav{\sigma^{M}}\geq \av{I} , \label{mepslaw}
\eeq
where $\aav{\sigma^{S}}$ and $\aav{\sigma^{M}}$ can be evaluated as follows:
\beqa
\aav{\sigma^{S}}&=& -S(\rho^{S}_{\mathrm{i}}) -\sum_{k,z}p_{k} \bbra{\phi^{S}_{k}(z)}\rho^{S}_{\mathrm{f}}(k)\kket{\phi^{S}_{k}(z)} \ln p_{\mathrm{r}}^{S}(z|k) \nonumber \\
&=& -\sum_{k}p_{k}\Tr[\rho^{S}_{\mathrm{f}}(k)\ln \rho^{S}_{\mathrm{r}}(k)] -S(\rho^{S}_{\mathrm{i}}), \label{sepdef}
\eeqa
where $\rho_{\mathrm{r}}^{S}(k):=\sum_{z}p^{S}_{\mathrm{r}}(z|k)\kket{\phi^{S}_{k}(z)}\bbra{\phi^{S}_{k}(z)}$ is the reference state of the system depending on $k$, and
\beqa
\aav{\sigma^{M}}&=& - \sum_{k}p_{k} \ln p^{M}_{\mathrm{r}}(k) -S(\rho^{M}_{0})\nonumber \\
&=& -\Tr[\rho^{M}_{\mathrm{f}}\ln\rho^{M}_{\mathrm{r}}]-S(\rho^{M}_{0}), 
\eeqa
with $\rho^{M}_{\mathrm{f}}:=\sum_{k}p_{k}\kket{\phi^{M}_{k}}\bbra{\phi^{M}_{k}}$ and $\rho^{M}_{\mathrm{r}}:=\sum_{k}p^{M}_{\mathrm{r}}(k)\kket{\phi^{M}_{k}}\bbra{\phi^{M}_{k}}$, being the final and reference states of the memory, respectively. We also note that $\av{I}$ in~(\ref{sepslaw}) and (\ref{mepslaw}) can be evaluated as follows:
\beqa
\av{I}&=&-\sum_{x}p^{S}_{\mathrm{i}}(x)\ln p^{S}_{\mathrm{i}}(x) + \sum_{k,y}p_{k}p(y|k) \ln p(y|k) \nonumber \\
&=& S(\rho^{S}_{\mathrm{i}})-\sum_{k}p_{k}S(\rho^{S}(k)) . \label{mutualdef}
\eeqa
The last equality indicates that $\av{I}$ gives the difference in the von Neumann entropy between the pre-measurement and postmeasurement states, which is the information gain~\cite{Groenewold,Ozawa} with inefficient measurements~\cite{Jacobs}. In general, $\av{I}$ can take on negative values by the following reason. Since there is a measurement back action in quantum systems, the entropy of the system increases, resulting in a higher von Neumann entropy compared with the premeasurement state. If the measurement back action is stronger than the acquired knowledge of the system due to the measurement, we end up with a negative information gain. For a classical system, the information gain is expressed by the classical mutual information content, which is nonnegative: $I_{\text{classical}}:=H(X)-\sum_{y}H(X|y)\geq 0$, where $H(X)$ is the Shannon entropy of the system and $H(X|y)$ is the Shannon entropy of the system conditioned on the outcome of the memory $y$. The difference from a quantum system arises from the fact that the indeterminacy of the initial state of the system does not deteriorate under a classical measurement since there is no back action.

Let us divide the information gain into two parts, the information gain $I_{\text{gain}}$ due to the measurement and the information loss $I_{\text{loss}}$ due to the measurement back action as $\av{I}=I_{\text{gain}}-I_{\text{loss}}$, where each term is defined as follows:
\beqa
& &I_{\text{gain}}:=S(\rho^{S}_{\mathrm{i}})-\sum_{k,a}p_{k,a}S(\rho^{S}(k,a))\geq 0, \label{igain}\\
& &I_{\text{loss}}:=\sum_{k}p_{k}S(\rho^{S}(k))-\sum_{k,a}p_{k,a}S(\rho^{S}(k,a))\geq 0, \label{iloss}
\eeqa
where
\beq
\rho^{S}(k,a):=\frac{M^{S}_{k,a}\rho^{S}_{\mathrm{i}}M^{\dagger S}_{k,a}}{p_{k,a}}
\eeq
is the postmeasurement state of the system when we know the label $a$ of the memory, and $p_{k,a}:=\Tr[M^{S}_{k,a}\rho^{S}_{\mathrm{i}}M^{\dagger S}_{k,a}]$. Both~(\ref{igain}) and (\ref{iloss}) are nonnegative since $I_{\text{gain}}$ is the information gain with the efficient measurement $M^{S}_{k,a}$~\cite{Ozawa}, and $I_{\text{loss}}$ is nonnegative from the inequality $S(\sum_{x}p(x)\rho_{x})\geq\sum_{x}p(x)S(\rho_{x})$~\cite{Nielsen} by noting that $\rho^{S}(k)=\sum_{a}p^{-1}_{k}p_{k,a}\rho^{S}(k,a)$. Here $I_{\text{gain}}$ represents the information gain due to the measurement when we know which initial state the memory is in. On the other hand, $I_{\text{loss}}$ measures the difference in the uncertainty of the postmeasurement state of the system with and without knowing which initial state the memory is in.

Let us examine a few examples: 

1. Suppose that the initial state of the memory is a pure state, i.e., $\rho^{M}_{0}=\ket{0}\bra{0}$. Then, the measurement operator $M^{S}_{k,a}$ reduces to $M^{S}_{k}$, which is called an efficient measurement~\cite{Ozawa,Jacobs}. In this case,  $\rho^{S}(k)=\rho^{S}(k,a)$ holds and the information loss is zero: $I_{\text{loss}}=0$, which means that there is no loss of information in estimating the postmeasurement state. This is the case for which $\av{I}\geq 0$. We note that Eq.~(\ref{QCfluc}) can be derived under this assumption, and thus we can take $\av{I}$ to be nonnegative in Eqs.~(\ref{QCfluc}) and (\ref{sepslaw}).

2. Suppose that the initial state of the system is a pure state, i.e., $\rho^{S}_{\mathrm{i}}=\kket{0}\bbra{0}$. Since $\rho^{S}(k,a)=p^{-1}_{k,a}M^{S}_{k,a}\kket{0}\bbra{0}M^{\dagger S}_{k,a}$ is also a pure state, we have $I_{\text{gain}}=0$, but $I_{\text{loss}}$ is nonzero in general since the postmeasurement state $\rho^{S}(k)$ is a mixed state due to the measurement back action and the initial uncertainty of the memory. This is the case for which $\av{I}\leq 0$. 

3. Suppose that both the initial state of the system and that of the memory are pure states. In this case, $\av{I}=I_{\text{gain}}=I_{\text{loss}}=0$, where we have neither the information gain nor loss.

Information processing allows us to reduce the entropy of the system up to the information gain as shown in~(\ref{sepslaw}), at the cost of increasing the entropy of the memory by the same quantity as shown in~(\ref{mepslaw}). From these inequalities, we find that $\aav{\sigma^{S}}+\aav{I}$ and $\aav{\sigma^{M}}-\aav{I}$ are nonnegative and express the true entropy productions, which measure the irreverisibility of the thermodynamic process for the system and the memory. Since the information exchange is made between the system and the memory, information gain~(\ref{genejarqcmutualdef}) cancels out and appears in neither the second law~(\ref{totalslaw}) nor the fluctuation theorem for the composite system, where the latter can be evaluated as follows:
\beqa
\av{\ee^{-\sigma^{SM}}}&=&\sum_{x,a,k,y,z}p(k,y|x,a)p^{M}_{\mathrm{r}}(k)p^{S}_{\mathrm{r}}(z|k)p(z|k,y)\nonumber \\
&=&\sum_{k,y,z}p^{M}_{\mathrm{r}}(k)p^{S}_{\mathrm{r}}(z|k)p(z|k,y)=1,
\eeqa
where the relation $\sum_{z}p^{S}_{\mathrm{r}}(z|k)=1$ and Eq.~(\ref{doublystochasticz}) are used in deriving the last equality.

If the assumption Eq.~(\ref{assumption}) does not hold, the left-hand side of Eq.~(\ref{QCfluc}) is rewritten as
\beq
\av{\ee^{-\sigma^{S}-I}}=\sum_{k,z,y\in Y} p_{k}p_{\mathrm{r}}(z|k)p(z|k,y),\eeq
which is not equal to unity in general. Here $Y$ is the set of labels $y$ satisfying $p(y|k)\neq 0$. A similar assumption that the conditional probability $P[y|\Gamma_{m}]$ of obtaining the measurement outcome $y$ conditioned on the state of the system being $\Gamma_{m}$ is always nonvanishing is needed to derive the integral fluctuation theorem under feedback control in Ref.~\cite{SagawaJar}, as pointed out in Ref.~\cite{Morikuni}. However, we can derive inequalities~(\ref{sepslaw}) and (\ref{mepslaw}) without assuming~(\ref{assumption}) by using Eqs.~(\ref{sepdef}) and (\ref{mutualdef}):
\beqa
\av{\sigma^{S}+I}&=&-\sum_{k}p_{k}\left[ S(\rho^{S}(k))+\Tr[\rho^{S}_{\mathrm{f}}(k)\ln\rho^{S}_{\mathrm{r}}(k)]\right] \nonumber \\
&=&-\sum_{k}p_{k}\left[S(\rho^{S}_{\mathrm{f}}(k))+\Tr[\rho^{S}_{\mathrm{f}}(k)\ln \rho^{S}_{\mathrm{r}}(k)]\right] \nonumber \\
&\geq& 0, \label{assumseclaw}
\eeqa
where we have used the fact that the von Neumann entropy does not change under a unitary transformation, and the positivity of the relative entropy~\cite{Nielsen}, i.e., $S(\rho||\sigma):=-S(\rho)-\Tr[\rho\ln\sigma]\geq 0$. 

If the assumption~(\ref{assumptiona}) is not met, the left-hand side of Eq.~(\ref{geneflucmemory}) is rewritten as
\beq
\aav{\ee^{-\sigma^{M}+I}}=\sum_{x\in X,a\in A,y\in Y,k}p^{M}_{\mathrm{r}}(k)p(y|k)p(k,y|x,a),
\eeq
where $p^{S}_{\mathrm{i}}(x)\neq 0$ and $p^{M}_{0}(a)\neq 0$ for $x\in X, a\in A$. We note that the inequality~(\ref{mepslaw}) also holds without the assumption~(\ref{assumptiona}) which can be shown in a similar manner as in Eq.~(\ref{assumseclaw}).

\subsection{\label{sec:QJE}Quantum Jarzynski equality}

In this subsection, we derive the quantum Jarzynski equality for the feedback-controlled system by assuming that the initial and reference states are given by canonical distributions: $\rho^{S}_{\mathrm{i}}=\ee^{-\beta (H^{S}_{\mathrm{i}}-F^{S}_{\mathrm{i}})}$ and $\rho^{S}_{\mathrm{r}}(k)=\ee^{-\beta(H^{S}_{\mathrm{f}}(k)-F^{S}_{k})}$, respectively, where $H^{S}_{\mathrm{i}}$ and $H^{S}_{\mathrm{f}}(k)$ are the initial and final Hamiltonians of the system. Then, the orthogonal bases $\{\kket{\psi^{S}(x)}\}$ and $\{\kket{\phi^{S}_{k}(z)}\}$ are given by the set of energy eigenfunctions: $H^{S}_{\mathrm{i}}\kket{\psi^{S}(x)}=E^{S}_{\mathrm{i}}(x)\kket{\psi^{S}(x)}$ and $H^{S}_{\mathrm{f}}(k)\kket{\phi^{S}_{k}(z)}=E^{S}_{\mathrm{f},k}(z)\kket{\phi^{S}_{k}(z)}$. Now $\sigma^{S}$ is related to the work done by the system as follows:
\beqa
\sigma^{S}(x,k,y)= -\beta\left[W^{S}(x,k,z)+\Delta f^{S}(k)\right], \label{entropywork}
\eeqa
where $W^{S}(x,k,z):=E^{S}_{\mathrm{i}}(x)-E_{\mathrm{f},k}^{S}(z)$ is the work done by the system, and $\Delta f^{S}(k):=F^{S}_{k}-F^{S}_{\mathrm{i}}$ is the free-energy difference. 
We now derive the following quantum Jarzynski equality under measurement and feedback by using Eq.~(\ref{QCfluc}):
\beq
\av{\ee^{\beta(W^{S}+\Delta f^{S})-I}}=1. \label{QCGeneJar}
\eeq
The convexity of the exponential functional and Eq.~(\ref{QCGeneJar}) reproduces the generalized second law under feedback control~\cite{Sagawa1}:
\beq
\aav{W^{S}}\leq -\aav{\Delta f^{S}} +\bz T\av{I},\label{genejarseclaw}
\eeq
where $\aav{W^{S}}= \Tr[\rho^{S}_{\mathrm{i}}H^{S}_{\mathrm{i}}]-\sum_{k}p_{k}\Tr[\rho^{S}_{\mathrm{f}}(k)H^{S}_{\mathrm{f}}(k)]$ is the averaged work done by the system. We can extract work from the system beyond the conventional second law, which is due to the correlation between the system and the memory.

To derive the quantum Jarzynski equality for the memory, we need to assume that the initial and reference states for the memory are given by canonical distributions. Then, we can formally derive the Jarzynski equality by using Eq.~(\ref{geneflucmemory}). However, we cannot transfer the entropy from the system to the memory via measurement and feedback since the initial state of the memory is given by a canonical distribution, resulting in a high entropy state. Since the Jarzynski equality for the memory does not hold if the initial state of the memory is a pure state, we will not discuss such a case in this paper. 

We note that the quantum fluctuation theorems under measurement and feedback control~(\ref{QCfluc}) reproduce the generalized second law~(\ref{genejarseclaw}) obtained in Ref.~\cite{Sagawa1}. Therefore, we may regard Eqs.~(\ref{QCfluc}),  (\ref{geneflucmemory}) and (\ref{QCGeneJar}) as the quantum extensions of the classical counterparts discussed in Ref.~\cite{Sagawa3}. The difference between our result and the one obtained in Ref.~\cite{Sagawa3} arises from the back action of the measurement, which is an essential feature in quantum systems. Moreover, the information gain $\av{I}$ can be negative due to the measurement back action. Our results include the classical case if conditional probabilities such as $p(y|x,k)$ and $p(z|k,y)$ are given by stochastic processes. In that case, the fluctuation theorems~(\ref{QCfluc}) and (\ref{geneflucmemory}) are obtained, with the measurement back action taken into account even for classical systems.

\section{\label{sec:summary}Conclusion}
We have derived quantum fluctuation theorems for both the system and the memory under measurements and feedback control. The obtained equalities include the information gain that expresses the acquired knowledge of the system due to the measurement. We also divide the information gain into two parts, namely the information gain due to the measurement and the information loss due to the lack of knowledge of the initial state of the memory. Both of these quantities are shown to be nonnegative. These equalities offer fundamental relations for controlling small fluctuating quantum systems even far from equilibrium.

This work was supported by Grants-in-Aid for Scientific Research (Kakenhi Nos. 22340114, 22103005, 24840015, 25287098, and 254105), Global COE Program ``The Physical Sciences Frontier," and the Photon Frontier Netowork Program of MEXT of Japan.

\end{document}